\documentclass[%
 reprint,
  superscriptaddress,
 amsmath,amssymb,amsfonts,
 aps,
pra,
 floatfix,
 longbibliography
]{revtex4-2}
\usepackage{comment}
\usepackage[utf8]{inputenc}
\usepackage[pdftex]{graphicx} \graphicspath{{}}
\usepackage{float} \usepackage{color}
\usepackage[pdftex,colorlinks=true]{hyperref}
\hypersetup{
  colorlinks=true,
  linkcolor=blue,
  citecolor = blue,
  urlcolor=blue,
}

\usepackage{mathtools}

\DeclarePairedDelimiterX\braket[2]{\langle}{\rangle}{#1 \delimsize\vert #2}
\DeclarePairedDelimiterX\expval[3]{\langle}{\rangle}{#1 \delimsize\vert #2  \delimsize\vert #3}
\DeclarePairedDelimiterX\proj[2]{\delimsize\vert#1\rangle}{\langle#2\delimsize\vert}{ }
\usepackage{siunitx}

\usepackage{tikz}
\definecolor{lime}{HTML}{A6CE39}
\DeclareRobustCommand{\orcidicon}{%
    \begin{tikzpicture}
    \draw[lime, fill=lime] (0,0) 
    circle [radius=0.13] 
    node[white] {{\fontfamily{qag}\selectfont\tiny ID}};
    \draw[white, fill=white] (-0.0625,0.095) 
    circle [radius=0.007];
    \end{tikzpicture}
    \hspace{-2mm}}

\newcommand{\orcidGD}{\href{https://orcid.org/0000-0002-7032-2465}{\orcidicon}}
\newcommand{\orcidFM}{\href{https://orcid.org/0000-0003-0506-0793}{\orcidicon}}
\newcommand{\orcidUD}{\href{https://orcid.org/0000-0002-5172-7352}{\orcidicon}}
\newcommand{\orcidCR}{\href{https://orcid.org/0009-0004-6708-076X}{\orcidicon}}

\begin{document}

\title{Pair localization in one-dimensional quasicrystals with spin-dependent hopping}
\author{Carlos Rosas\orcidCR}
\author{U. A. D\'iaz-Reynoso\orcidUD}
\author{Francisco Mireles\orcidFM}
\author{G. A. Dom\'inguez-Castro\orcidGD}
\email{gustavodomin@ens.cnyn.unam.mx}
\affiliation{Centro de Nanociencias y Nanotecnología, Universidad Nacional Autónoma de México, Apartado Postal 14, 22800 Ensenada, Baja California, Mexico}
\date{\today}

\begin{abstract}
We investigate the quantum localization of two interacting particles moving in a one-dimensional quasicrystal with unequal hopping amplitudes. Using a two-body Green's function approach, we determine the localization phase diagram of attractively and repulsively bound pairs, showing that increasing the hopping amplitude of one constituent while keeping the other fixed raises the critical quasiperiodic strength, promoting pair delocalization. The localization threshold nevertheless remains bounded above by the Aubry–André critical point of the slower constituent. Effective analytical descriptions capture pair localization in the limits of strong hopping imbalance and strong interactions. The dynamics of an initially on-site pair provides signatures of delocalization induced by unequal hopping amplitudes. In particular, the effective spreading exponent 
of the root-mean-square width of a pair distribution reveals a progression from localized to subdiffusive and eventually superdiffusive behavior as the hopping ratio increases. Our results establish unequal hopping amplitudes as a means of tuning the transport properties of interacting pairs in quasiperiodic lattices, with direct relevance to ultracold atoms in optical lattices.
\end{abstract}

\maketitle

\section{Introduction}

Quasicrystals are intriguing structures that constitute an intermediate case between disordered and periodic crystals, exhibiting long-range order without spatial periodicity~\cite{PhysRevLett.53.2477}. Their unconventional structure provides a paradigmatic setting for exploring quantum localization. In particular, one-dimensional quasicrystals, commonly described by the celebrated Aubry-Andr\'e model~\cite{aubry1980analyticity}, have attracted considerable theoretical~\cite{PhysRevLett.114.146601,PhysRevA.81.013614,PhysRevLett.123.070405,PhysRevLett.126.106803,PhysRevB.105.L220201} and experimental~\cite{roati2008anderson,PhysRevLett.103.013901,PhysRevLett.120.160404} interest. In contrast to the one-dimensional Anderson model~\cite{PhysRev.109.1492}, where arbitrarily weak uncorrelated disorder localizes all single-particle states, quasiperiodic lattices exhibit a localization transition at a finite modulation strength, with critical states emerging at the transition point~\cite{wilkinson1984critical}. 

The inclusion of interparticle interactions at a finite density can substantially modify localization in quasicrystals. For sufficiently strong quasiperiodicity, thermalization can be suppressed, leading to a dynamical regime in which the system retains information of its initial state over long times~\cite{schreiber2015observation}. Regardless of considerable progress~\cite{PhysRevLett.113.045304,PhysRevLett.129.103401,PhysRevLett.115.180401,PhysRevLett.115.186601}, understanding how quasiperiodicity and interparticle interactions shape quantum localization in the many-body scenario remains challenging~\cite{PhysRevB.105.224203}. In this context, interacting few-particle disordered systems provide a crucial framework for identifying physical mechanisms that can help elucidate the behavior of complex many-body systems. In particular, despite its apparent simplicity, the pair localization problem already exhibits rich phenomenology, including interaction-induced enhancement of the pair localization length~\cite{PhysRevLett.73.2607,PhysRevLett.76.491,PhysRevB.100.224203}, multifractal spectrum~\cite{PhysRevB.101.174201}, mobility edges~\cite{PhysRevResearch.2.033501}, and peculiar dynamical regimes~\cite{PhysRevA.104.033306}.

Typically, pair localization is studied under the assumption that the constituents have equal effective masses and, consequently, identical tunneling amplitudes~\cite{PhysRevLett.109.155306,PhysRevB.106.134208,PhysRevB.102.144201,PhysRevB.54.14896,flach2012correlated,PhysRevA.89.023823}. However, composite quantum objects with constituents characterized by different effective masses arise naturally in several condensed-matter settings. A prominent example is provided by excitons in transition metal dichalcogenides, where electrons and holes originate from different conduction and valence bands and generally possess unequal effective masses~\cite{RevModPhys.90.021001,rivera2018interlayer,jiang2021interlayer}. Multiband superconductors provide another example in which paired quasiparticles may originate from bands characterized by substantially different effective masses~\cite{PhysRevB.110.094517}. Moreover, these composite particles can be realized in disordered media~\cite{58pr-k9vt}, motivating the study of how pair localization is modified in the presence of unequal kinetic scales.

Ultracold atoms trapped in optical lattices represent a powerful quantum simulation platform for realizing interacting disordered systems under highly controlled conditions. The ability to tune lattice geometry~\cite{lewenstein2007ultracold} and interaction strengths~\cite{RevModPhys.80.885}, together with the possibility of engineering tailored disordered potentials~\cite{sanchez2010disordered}, has enabled the experimental observation of quantum localization in quasicrystals from the single-particle regime~\cite{roati2008anderson} to the interacting many-body limit~\cite{schreiber2015observation}. Recent advances in species-dependent~\cite{PhysRevA.75.053612,PhysRevA.77.011603} and state-dependent~\cite{PhysRevLett.91.010407,PhysRevLett.124.203201} optical potentials allow different atomic species or internal states to experience distinct lattice depths and, consequently, different tunneling amplitudes, providing a natural setting for exploring pair localization in the presence of multiple kinetic energy scales. Furthermore, quantum gas microscopy enables direct access to the particle dynamics through single-site resolution of individual atoms~\cite{gross2021quantum,sherson2010single,bakr2009quantum}, offering a direct route to probe localization at the few-particle level.

In this work, we investigate how unequal hopping amplitudes shape pair-localization and pair-transport of two interacting particles in a one-dimensional quasicrystal. Using a two-body Green’s-function approach, we determine the localization phase diagram of the attractively bound dimer
ground state and the highest-energy repulsively bound pair, showing that the critical quasiperiodic strength crucially depends on the hopping ratio of the pair constituents. In particular, increasing the tunneling amplitude of one particle while keeping the other fixed raises the critical quasiperiodic strength, promoting pair delocalization. The localization threshold is nevertheless bounded  from above by the Aubry–André critical point of the slower constituent. We provide effective analytical descriptions that accurately capture pair localization in the limits of strong hopping imbalance and strong interactions. Within the localized regime, we numerically demonstrate that the hopping ratio allows continuous tuning of the pair localization length. We further investigate dynamical signatures of pair delocalization by tracking the time evolution of an initially on-site pair. From the effective spreading exponent of the root-mean-square width of the normalized on-site pair distribution, we identify the progression from localized to subdiffusive and eventually superdiffusive dynamics as the hopping ratio increases.

\section{Model}
We consider two distinguishable interacting particles, labeled by a pseudospin index $\sigma=\uparrow,\downarrow$, moving in a one-dimensional quasicrystal with unequal hops and lattice constant $a$, set to $a=1$ below. The Hamiltonian of the two-body system can be written as $\hat{H}=\hat{H}_{0}+\hat{U}$, with $\hat{H}_{0}$ the noninteracting component and $\hat{U}$ denoting the particle-particle interaction operator. The noninteracting part of $\hat{H}$ can be decomposed as $\hat{H}_{0} = \hat{H}_{\uparrow}\otimes\mathbb{I}_{\downarrow} + \mathbb{I}_{\uparrow}\otimes\hat{H}_{\downarrow}$, where $\mathbb{I}_{\sigma}$ denotes the identity operator in the Hilbert space of particle $\sigma$, and $\hat{H}_{\sigma}$ is the single-particle Aubry-Andr\'e (AA) Hamiltonian
\begin{equation}
\begin{split}
\hat{H}_{\sigma} &= -J_{\sigma}\sum_{j}(\hat{c}_{j+1,\sigma}^{\dagger}\hat{c}_{j,\sigma}+h.c.)\\
&+\Delta\sum_{j}\cos(2\pi\beta j + \phi)\hat{n}_{j,\sigma}.
\end{split}
\label{Eq1}
\end{equation}
Here, $\hat{c}_{j,\sigma}^{\dagger}$ $(\hat{c}_{j,\sigma})$ creates (annihilates) a particle of component $\sigma$ at lattice site $j$, and $\hat{n}_{j,\sigma}=\hat{c}_{j,\sigma}^{\dagger}\hat{c}_{j,\sigma}$ is the corresponding number operator. The nearest-neighbor hopping rate for component $\sigma$ is $J_{\sigma}$. The quasiperiodic potential is characterized by its strength $\Delta$, the incommensurate parameter $\beta=(\sqrt{5}-1)/2$, and the phase offset $\phi\in[0,2\pi)$. For the single-particle problem, all eigenstates of the AA model are extended for $\Delta/J_{\sigma}<2$, exponentially localized for $\Delta/J_{\sigma}>2$, and critical at the self-dual point $\Delta/J_{\sigma}=2$ \cite{aubry1980analyticity,wilkinson1984critical,dominguez2019aubry}. Spin- or species-dependent optical lattices can yield different quasiperiodic potential amplitudes for the two particles. To isolate the effects of unequal hopping amplitudes, we consider equal quasiperiodic strengths, $\Delta_{\uparrow}=\Delta_{\downarrow}=\Delta$.

Interactions between the two particles are introduced via an on-site coupling  of strength $U$,
\begin{equation}
\hat{U} = U\sum_{j}\hat{n}_{j,\uparrow}\hat{n}_{j,\downarrow}.
\label{Eq2}
\end{equation}
Since the pair is confined to one dimension, the two-particle bound state exists for arbitrarily weak attractive ($U<0$) or repulsive ($U>0$) interaction strengths~\cite{valiente2008two,ding2023polarons}. For lattices with nearest-neighbor hopping, the wave functions of the attractively bound dimer ground state and the highest-energy repulsively bound dimer are related by a spectral symmetry~\cite{PhysRevLett.119.225302}. The corresponding states thus have identical localization properties~\cite{PhysRevLett.109.155306}. We therefore express results common to both cases in terms of the interaction magnitude $|U|$.

\section{Green's function formalism}
The two-body Schr\"odinger equation can be written as 
\begin{equation}    
(E-\hat{H}_{0})|\Psi\rangle = \hat{U}|\Psi\rangle.
\label{Eq3}
\end{equation}
This equation can be solved numerically using the noninteracting two-body Green’s function operator $\hat{\mathcal{G}}_{0}(E)=(E-\hat{H}_{0})^{-1}$, which can be formally written in terms of the eigenstates $|\varphi_{\ell,\sigma}\rangle$ and eigenenergies $\varepsilon_{\ell,\sigma}$ of the AA Hamiltonian
\begin{equation}
\hat{\mathcal{G}}_{0}(E) = \sum_{m,n}\frac{1}{E-\varepsilon_{n,\uparrow}-\varepsilon_{m,\downarrow}}|\varphi_{n,\uparrow}\varphi_{m,\downarrow}\rangle\langle\varphi_{n,\uparrow}\varphi_{m,\downarrow}|.
\label{Eq4}
\end{equation}

Applying $\hat{\mathcal{G}}_{0}$ to both sides of Eq.~\eqref{Eq3}, one can find $|\Psi\rangle=\hat{\mathcal{G}}_{0}\hat{U}|\Psi\rangle$. Projecting this expression over the state $|j_{\uparrow}, j_{\downarrow}'\rangle$, we obtain an equation for the amplitudes $\Psi (j_{\uparrow},j'_{\downarrow}) = \langle j_{\uparrow}, j_{\downarrow}'|\Psi\rangle$ of the two-particle wave function:
\begin{equation}
\begin{split}
\Psi(j_{\uparrow},j'_{\downarrow}) &= \langle j_{\uparrow}, j'_{\downarrow}|\hat{\mathcal{G}}_{0}\hat{U}|\Psi\rangle\\
&= U\sum_{i}\langle j_{\uparrow}, j'_{\downarrow}|\hat{\mathcal{G}}_{0}|i_{\uparrow},i_{\downarrow}\rangle\Psi(i_{\uparrow},i_{\downarrow}),
\end{split}
\label{Eq5}
\end{equation}
where in the last equality, we use the fact that the interaction operator $\hat{U}$ acts only when both particles occupy the same site. Equation~\eqref{Eq5} shows that the diagonal amplitudes of $\Psi(j_{\uparrow},j'_{\downarrow})$ contain all the independent wave function information needed to reconstruct the full two-body state. Thus, the original $L^2$-dimensional two-particle problem, where $L$ is the number of lattice sites, reduces to an $L$-dimensional energy-dependent eigenvalue problem. Setting $j_{\uparrow}=j'_{\downarrow}=j$ in Eq.~\eqref{Eq5} yields the desired eigenvalue problem:
\begin{equation}
\sum_{i}G_{0}(j,i;E)\psi(i) = \frac{1}{U}\psi(j),
\label{Eq6}
\end{equation}
with $\psi(j)=\Psi(j_{\uparrow},j_{\downarrow})$ and $G_{0}(j,i;E)$ the matrix elements of the noninteracting two-body Green's function operator:
\begin{equation}
\begin{split}
G_{0}(j,i;E) &= \langle j_{\uparrow}, j_{\downarrow}|\hat{\mathcal{G}}_{0}(E)|i_{\uparrow},i_{\downarrow}\rangle\\
&= \sum_{n,m} \frac{\varphi_{n,\uparrow}(j)\varphi_{m,\downarrow}(j)\varphi_{n,\uparrow}^{*}(i)\varphi_{m,\downarrow}^{*}(i)}{E-\varepsilon_{n,\uparrow}-\varepsilon_{m,\downarrow}}.
\end{split}
\label{Eq7}
\end{equation}
Equation~\eqref{Eq6} can be solved self-consistently~\cite{PhysRevLett.109.155306,PhysRevB.106.134208}. For attractive interactions ($U<0$), one starts by calculating the matrix $G_{0}(E)$ at a trial energy below the noninteracting two-particle spectrum, $E<\varepsilon_{0,\uparrow}+\varepsilon_{0,\downarrow}$, and computes its lowest eigenvalue $\lambda_{0}(E)$. The energy is then varied until $\lambda_{0}(E)=1/U$. The corresponding lowest eigenvector is the desired state. For the time-independent analysis, we consider a lattice of $L=987$ sites with open boundary conditions and average all reported quantities over $20$ realizations of the phase shift $\phi$, sampled uniformly from $[0,2\pi)$.
\begin{figure}[t!]
\centering
\includegraphics[width=0.99\columnwidth]{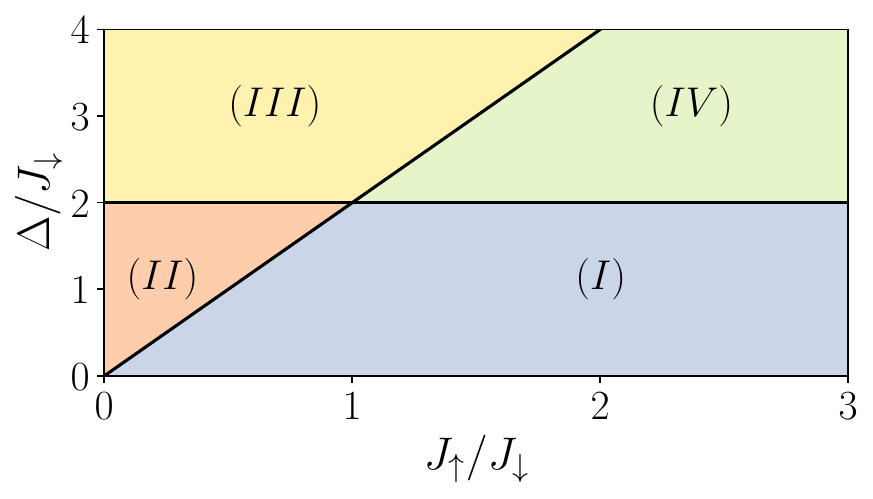}
\caption{Extended-localized regimes for two noninteracting particles with unequal hopping amplitudes. In region I, defined by $\Delta/J_{\downarrow}<2$ and $\Delta/J_{\downarrow}<2J_{\uparrow}/J_{\downarrow}$, both particles are extended. For region II, delimited by $2J_{\uparrow}/J_{\downarrow}<\Delta/J_{\downarrow}<2$, the $\downarrow$ particle is extended while the $\uparrow$ particle is localized. In region III, defined by $\Delta/J_{\downarrow}>2$ and $\Delta/J_{\downarrow}>2J_{\uparrow}/J_{\downarrow}$, both particles are localized. Lastly, region IV delimited by $2J_{\uparrow}/J_{\downarrow}>\Delta/J_{\downarrow}>2$, the $\downarrow$ particle is localized, whereas $\uparrow$ particle is extended.\label{Fig1}}
\end{figure}

\section{Ground State Pair Localization}
We begin by considering the localization problem for two noninteracting particles with spin-dependent hopping. In the absence of interactions, no bound state is formed, and the two-body localization properties follow from those of the individual constituents. Each particle undergoes the Aubry-Andr\'e transition at a quasiperiodic strength determined by its own tunneling amplitude. Unequal hopping amplitudes give rise to four distinct regimes in which the two particles can be simultaneously extended, simultaneously localized, or exhibit contrasting spatial behavior. These regimes are summarized in Fig.~\ref{Fig1} as a function of quasiperiodicity $\Delta/J_{\downarrow}$ and hopping ratio $J_{\uparrow}/J_{\downarrow}$.

\begin{figure}[t!]
\centering
\includegraphics[width=0.99\columnwidth]{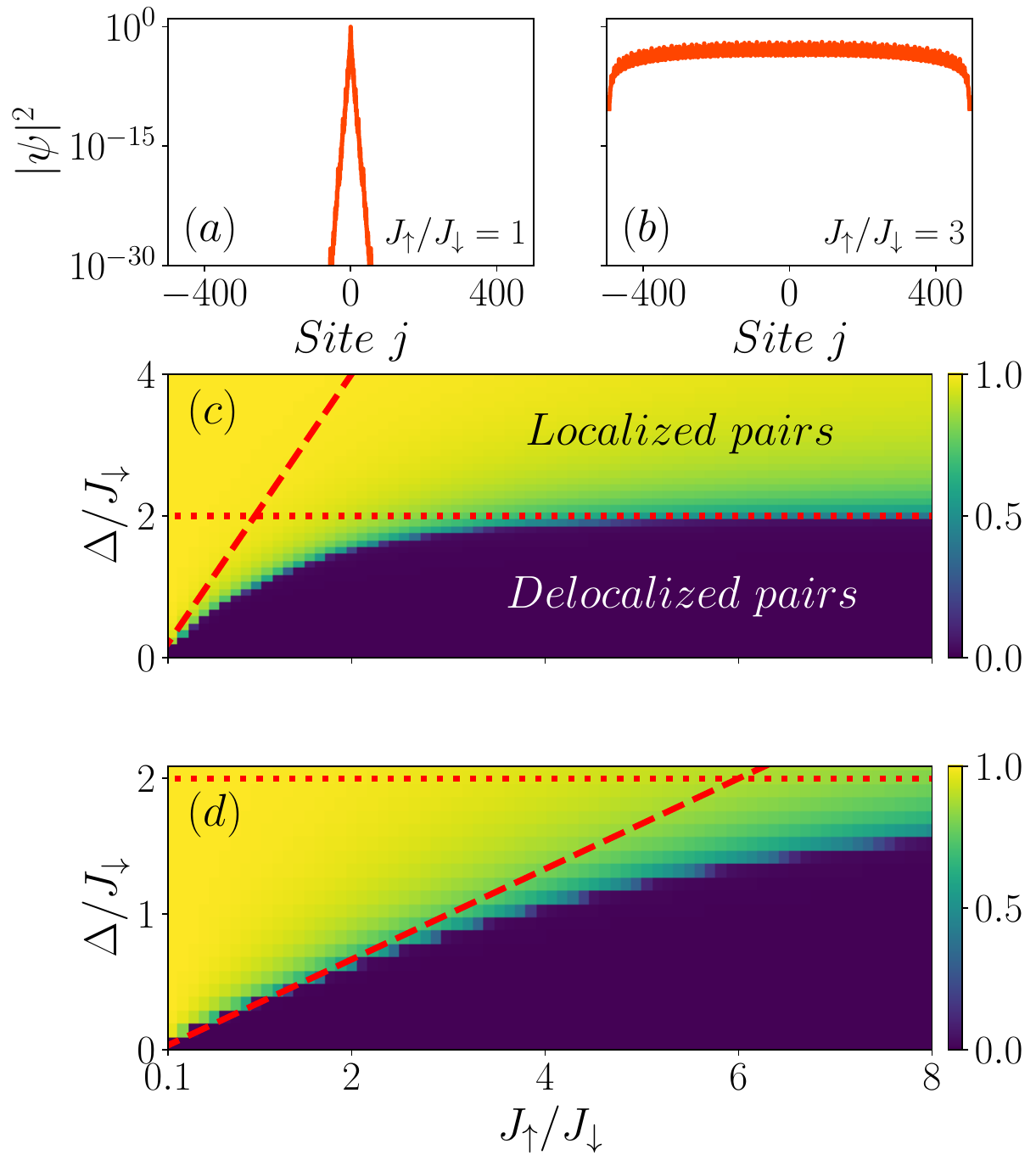}
\caption{(a)-(b) Bound pair density probability $|\psi|^2$ as a function of the lattice index $j$ for $J_{\uparrow}/J_{\downarrow}=1$ and $J_{\uparrow}/J_{\downarrow}=3$, respectively. Both panels consider $|U|/J_{\downarrow}=1$ and $\Delta/J_{\downarrow}=1.5$. (c)-(d) Dimer inverse participation ratio as a function of the hopping ratio $J_{\uparrow}/J_{\downarrow}$ and the quasiperiodic strength $\Delta/J_{\downarrow}$ for $|U|/J_{\downarrow}=1$ and $|U|/J_{\downarrow}=6$, respectively. The red dotted line corresponds to the AA critical point $\Delta_{c}/J_{\downarrow}=2$, 
whereas the red dashed line is associated with the strong-coupling prediction $\Delta_{SC}=2J_{\uparrow}J_{\downarrow}/|U|$
\label{Fig2}}
\end{figure}

Finite on-site interactions bind the two particles into a correlated dimer. As shown in Figs.~\ref{Fig2}(a)-\ref{Fig2}(b), unequal hops can substantially modify the spatial behavior of the dimer. For $\Delta/J_{\downarrow}=1.5$, the bound state associated with the identical tunneling case $J_{\uparrow}/J_{\downarrow}=1$ is localized. In contrast, increasing the hopping ratio $J_{\uparrow}/J_{\downarrow}=3$ results in an extended bound pair.

To characterize the extended-localized behavior of the pair, we employ the inverse participation ratio (IPR)
as a measure of localization. Given a normalized wave function $\psi$ its IPR is defined as IPR$_{\psi} =\sum_{i=1}^{L}|\psi(i)|^{4}$. For extended states, the IPR vanishes in the thermodynamic limit as $\propto L^{-1}$, whereas it remains finite for localized states. In Figs.~\ref{Fig2}(c)-\ref{Fig2}(d), we plot the inverse participation ratio as a function of $J_{\uparrow}/J_{\downarrow}$ and $\Delta/J_{\downarrow}$ for $|U|/J_{\downarrow}=1$ and $|U|/J_{\downarrow}=6$, respectively. Compared with the noninteracting diagram of Fig.~\ref{Fig1}, interactions substantially modify localization boundaries, the bound state is localized throughout regions II-IV, whereas an extended dimer survives within a fraction of region I. 

Interestingly, as $J_{\uparrow}/J_{\downarrow}$ increases beyond the identical hopping case ($J_{\uparrow}=J_{\downarrow}$), the pair localization threshold rises, promoting pair delocalization, and approaches the AA critical point of the slower constituent, $\Delta_c/J_{\downarrow}=2$. This behavior can be understood from an effective description of the dimer in the regime of strong hopping imbalance ($J_{\uparrow}\gg J_{\downarrow}$), where 
the tunneling and quasiperiodic amplitudes are given by (see Appendix~\ref{Quasiperiodic})
\begin{equation}
\begin{split}
J_{dimer} &\simeq \frac{2J_{\uparrow}J_{\downarrow}}{\sqrt{|U|^2+4J_{\uparrow}^{2}}},\\ 
\Delta_{dimer} &\simeq\Delta\left[1+\frac{|U|^{2}}{|U|^{2}+4J_{\uparrow}^{2}\sin^{2}(\pi\beta)}\right].
\end{split}
\label{Eq8}
\end{equation}
For $J_{\uparrow}\gg J_{\downarrow},|U|,\Delta$, the effective hopping approaches $J_{\mathrm{dimer}}\to J_{\downarrow}$, indicating that increasing $J_\uparrow$ no longer enhances the mobility of the bound state and that the pair motion is limited by the slower constituent. Meanwhile, the increasing kinetic scale of the faster constituent suppresses its contribution to the effective quasiperiodic modulation through spatial averaging. Consequently, the pair experiences an effective potential whose amplitude approaches $\Delta_{dimer}\rightarrow\Delta$, as $J_{\uparrow}\gg J_{\downarrow},|U|,\Delta$. The effective AA localization criterion therefore yields $\Delta_{dimer,c}/J_{\downarrow}\to2$.

As shown in Fig.~\ref{Fig2}(d), increasing the interaction strength  further suppresses the extended region. To provide analytical insight into dimer localization within the strongly interacting regime $|U|\gg J_{\uparrow},J_{\downarrow},\Delta$, we project the full two-body Hamiltonian within the subspace of on-site dimers $|j,j\rangle$. To second order in the single-particle hopping, the resulting effective Hamiltonian is
\begin{equation}
\begin{split}
\hat{H}_{dimer} &= -\frac{2J_{\uparrow}J_{\downarrow}}{|U|}\sum_{j}(\hat{d}_{j+1}^{\dagger}\hat{d}_{j}+h.c.)\\ &+ 2\Delta\sum_{j}\cos(2\pi\beta j +\phi)\hat{n}_{j,d},
\end{split}
\label{Eq9}
\end{equation}
where $\hat{d}_{j}^{\dagger}=\hat{c}_{j,\uparrow}^{\dagger}\hat{c}_{j,\downarrow}^{\dagger}$ ($\hat{d}_{j}=\hat{c}_{j,\downarrow}\hat{c}_{j,\uparrow}$) is the dimer creation (annihilation) operator and $\hat{n}_{j,d} = \hat{n}_{j,\uparrow}\hat{n}_{j,\downarrow}$ the corresponding number operator. The effective hopping of the dimer is $J_{dimer}=2J_{\uparrow}J_{\downarrow}/|U|$, whereas the quasiperiodic strength is $\Delta_{dimer}=2\Delta$. Notice that the effective amplitudes in Eq.~\eqref{Eq8} account for both dimer hopping and quasiperiodic modulation in the strong-coupling regime, $|U|\gg J_{\uparrow},J_{\downarrow},\Delta$, highlighting the significance of the approach. The dimer Hamiltonian in Eq.~\eqref{Eq9} has the form of a single-particle Aubry--André model.  Applying the corresponding localization criterion $\Delta_{c}/J=2$, gives $\Delta_{dimer,c}=2J_{\uparrow}J_{\downarrow}/|U|$. The strong-coupling prediction, shown by the red dashed lines in Figs.~\ref{Fig2}(c) and \ref{Fig2}(d), reproduces the numerical localization boundary for $|U|\gg J_{\uparrow},J_{\downarrow},\Delta$. Deviations emerge when $J_{\uparrow}\sim |U|$, where the tightly bound on-site dimer approximation breaks down, demanding a full two-body Green's function approach.
\begin{figure}[t!]
\centering
\includegraphics[width=0.99\columnwidth]{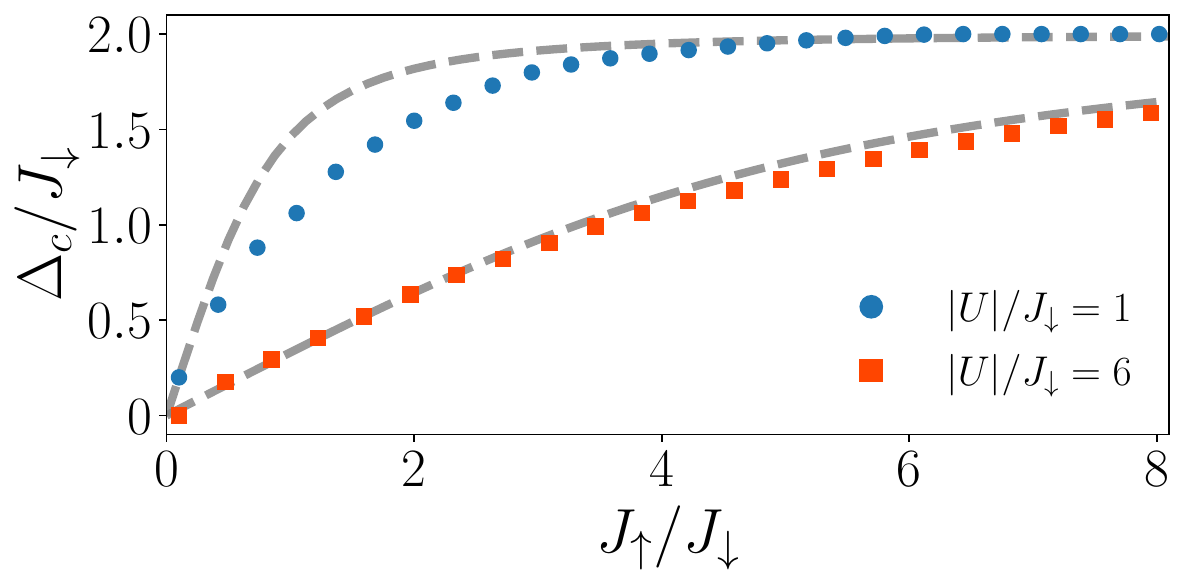}
\caption{Critical quasiperiodic strength $\Delta_c/J_\downarrow$ as a function of the hopping ratio $J_\uparrow/J_\downarrow$ for two different interaction strengths $|U|/J_{\downarrow}$. Markers denote numerical results, gray dashed lines show the effective dimer prediction in Eq.~\eqref{Eq10}.\label{Figure3}}
\end{figure}

To determine the critical quasiperiodic modulation $\Delta_{c}$ associated with the pair localization transition, we compute the inflection point of the inverse participation ratio. That is, for fixed $U/J_{\downarrow}$ and $J_{\uparrow}/J_{\downarrow}$, we calculate the IPR as a function of $\Delta/J_{\downarrow}$ and identify $\Delta_{c}$ at the point where the second derivative vanishes. The inflection point technique has been successfully employed in previous studies, yielding consistent results with alternative approaches, such as the inverse and normalized participation ratio crossing~\cite{PhysRevB.106.134208,dominguez2019aubry}. Figure~\ref{Figure3} illustrates the resulting $\Delta_{c}$ as a function of the hopping ratio $J_{\uparrow}/J_{\downarrow}$ for two interaction strengths. Markers correspond to numerical results, whereas the gray dashed lines are associated with the effective dimer prediction 
\begin{equation}
\Delta_{dimer,c} \simeq \frac{4J_{\uparrow}J_{\downarrow}}{\sqrt{|U|^2+4J_{\uparrow}^{2}}}\left(1+\frac{|U|^{2}}{|U|^{2}+4J_{\uparrow}^{2}\sin^{2}(\pi\beta)}\right)^{-1}.
\label{Eq10}
\end{equation}
The analytical expression in Eq.~\eqref{Eq10} 
accurately interpolates between the strong-coupling and the strong hopping imbalance limits, while capturing the pair localization transition reasonably well in the intermediate regime $J_{\uparrow}\sim J_{\downarrow}\sim |U|$. Moreover, it also quantifies the approach to the asymptotic localization threshold. 
\begin{figure}[t!]
\centering
\includegraphics[width=0.99\columnwidth]{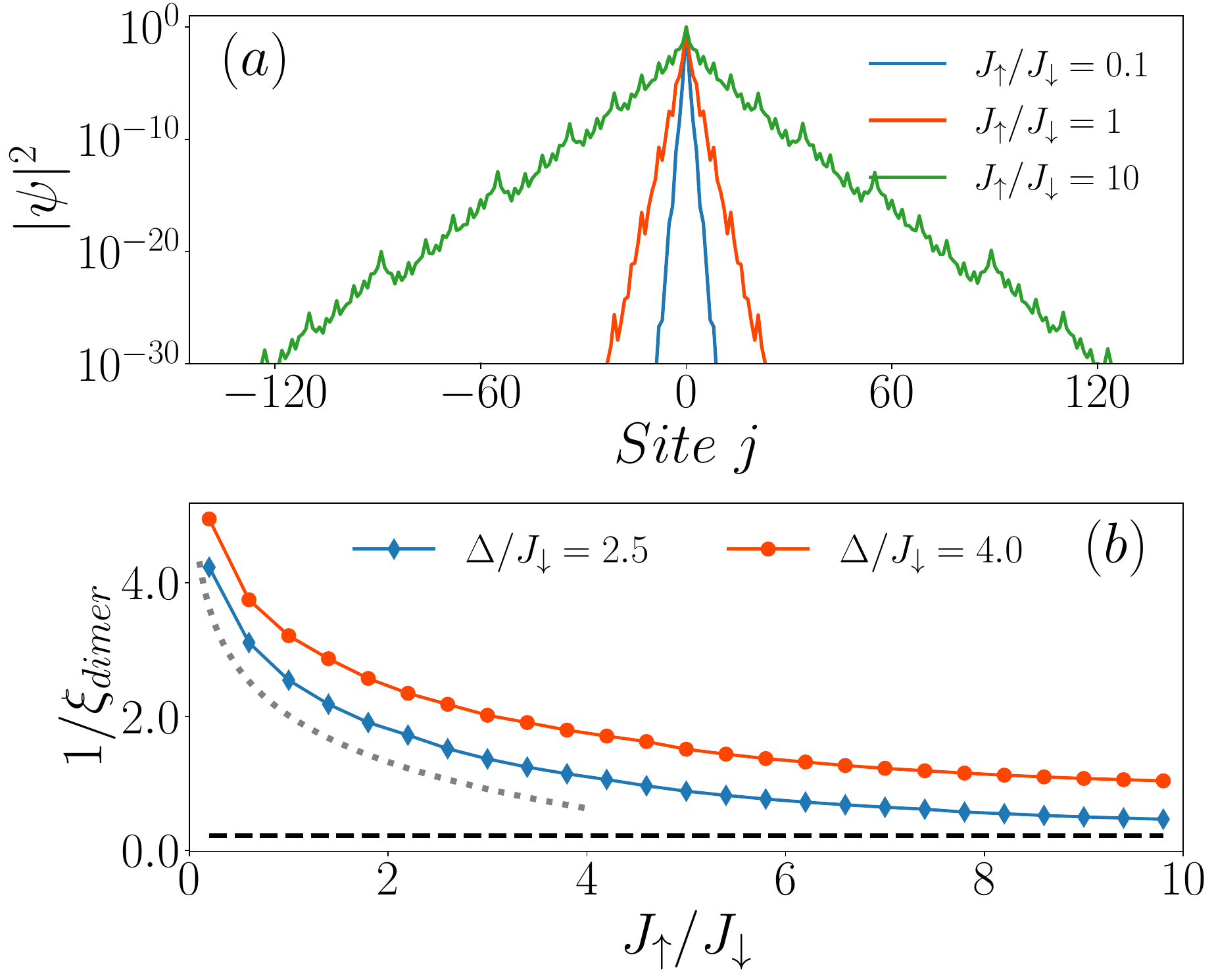}
\caption{(a) Bound-pair probability, shown on a logarithm scale as a function of the lattice index $j$ for several hopping ratios at a fixed interaction strength of $|U|/J_{\downarrow}=1$ and quasiperiodic strength $\Delta/J_{\downarrow}=2.5$. (b) Inverse localization length $1/\xi_{dimer}$ as a function of $J_{\uparrow}/J_{\downarrow}$ for two quasiperiodic strengths. 
The horizontal dashed line indicates the corresponding single-particle result for the slower component, $1/\xi_{\downarrow}=\ln \Delta/2J_{\downarrow}$ at $\Delta/J_{\downarrow}=2.5$, while the gray dotted curve displays the strong-coupling prediction $1/\xi_{SC}=\ln \Delta|U|/(2J_{\downarrow}J_{\uparrow})$ at $(\Delta/J_{\downarrow},|U|/J_{\downarrow})=(2.5,6)$. \label{Figure4}}
\end{figure}

As it is well known, localization in quasiperiodic lattices with
nearest-neighbor hops is characterized  by exponentially decaying spatial profiles $|\psi|\sim e^{-|j-j_{0}|/\xi_{dimer}}$, where $\xi_{dimer}$ denotes the localization length and $j_{0}$ the localization center. Figure~\ref{Figure4}(a) shows the dimer probability $|\psi|^{2}$ for several hopping ratios at fixed quasiperiodic and interaction strengths $(\Delta/J_{\downarrow},|U|/J_{\downarrow})=(2.5,1)$. As can be seen, the spatial decay of the bound state crucially depends on the hopping ratio, resulting in a larger localization length as $J_{\uparrow}/J_{\downarrow}$ increases.

To quantify the role of $J_{\uparrow}/J_{\downarrow}$ on $\xi_{dimer}$, we extract the localization length by fitting the tails of the probability amplitude $|\psi|$ to the exponential form $e^{-|j-j_{0}|/\xi_{dimer}}$. Figure~\ref{Figure4}(b) illustrates the inverse localization length $1/\xi_{dimer}$ as a function of $J_{\uparrow}/J_{\downarrow}$ for $|U|/J_{\downarrow}=6$. As can be seen, $1/\xi_{dimer}$ decreases monotonically, thus $\xi_{dimer}$ increases, with the hopping ratio, providing a direct real-space signature of pair delocalization induced by unequal hopping amplitudes. For small to moderate hopping ratios, $1/\xi_{dimer}$ is bounded from below by the strong-coupling prediction, $1/\xi_{SC}=\ln\Delta|U|/(2J_{\downarrow}J_{\uparrow})$. As  $J_{\uparrow}/J_{\downarrow}$ increases beyond the identical
hopping case, $1/\xi_{dimer}$ approaches the single-particle value of the slower constituent, $1/\xi_{\downarrow} =\ln\Delta/2J_{\downarrow}$, consistent with the effective dimer description in the limit of strong hopping imbalance.

\begin{figure}[t!]
\centering
\includegraphics[width=0.99\columnwidth]{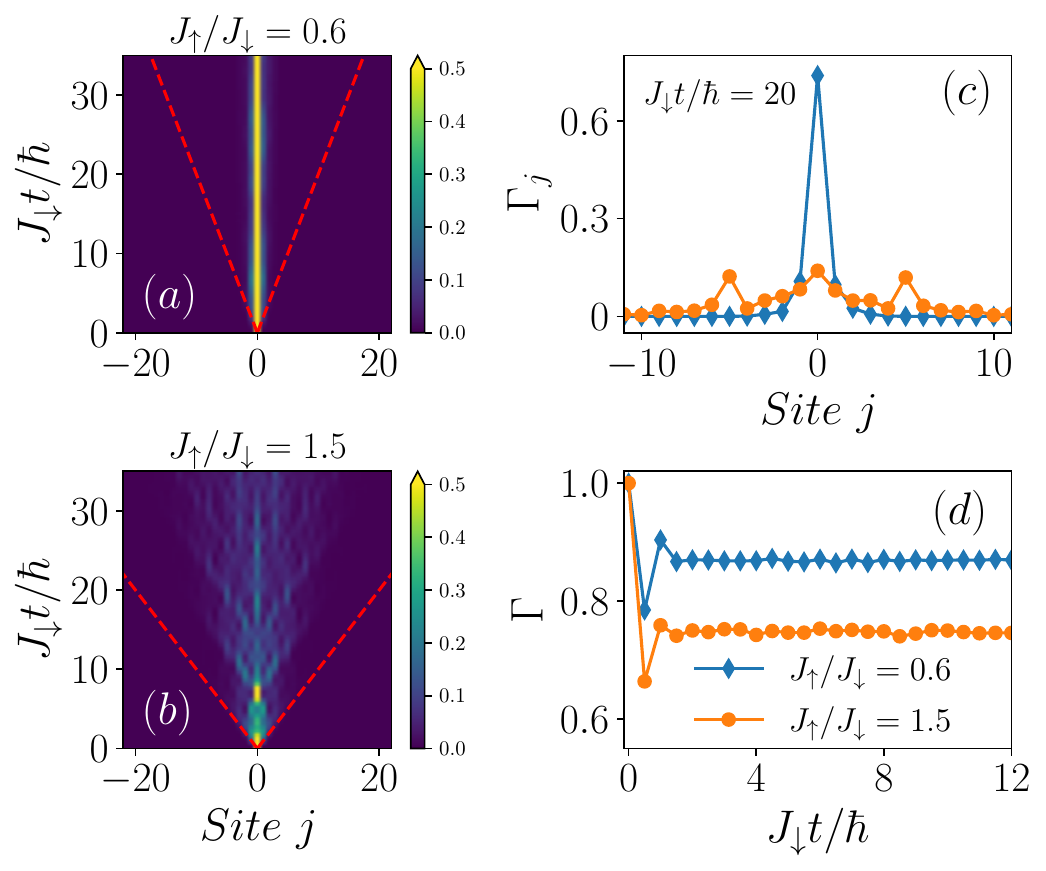}
\caption{(a)-(b) Time evolution of the normalized on-site distribution for $J_{\uparrow}/J_{\downarrow}=0.6$ and $1.5$, respectively. We consider $(\Delta/J_{\downarrow},U/J_{\downarrow})=(0.4,-6)$. The red dashed lines in each panel correspond to the ballistic fronts predicted for a tightly bound pair in the absence of quasiperiodicity $\pm 4J_{\uparrow}J_{\downarrow}t/(\hbar|U|)$. (c) Corresponding profiles of $\Gamma_{j}$ at $J_{\downarrow}t/\hbar=20$. (d) Time evolution of the total probability of finding both particles on the same site for $(\Delta/J_{\downarrow},U/J_{\downarrow})=(0.4,-6)$. Results are obtained for a lattice of $L=101$ sites and averaged over $50$ random realizations of the phase $\phi$.\label{Fig5}}
\end{figure}

\section{Pair Dynamics}

In this section, we investigate dynamical signatures of pair delocalization induced by unequal hopping amplitudes. For this purpose, we consider two particles initially occupying the central lattice site, $|\Psi(0)\rangle=
\hat c_{L/2,\uparrow}^{\dagger}
\hat c_{L/2,\downarrow}^{\dagger}|0\rangle
$. This configuration provides a natural initial state for probing pair dynamics in ultracold-atom experiments~\cite{kwan2024realization,winkler2006repulsively}. For sufficiently strong interactions, $|\Psi(0)\rangle$ has significant overlap with the two-body bound state sector, allowing correlated pair propagation to be resolved. However, because dimer wave functions extend over finite interparticle separations, the initial on-site configuration also retains a finite overlap with scattering states. Consequently, the resulting dynamics contains contributions from both bound-pair motion and unbound-particle propagation. 

The time evolution of the initial state is obtained by exact diagonalization of the full Hamiltonian $\hat{H}=\hat{H}_{0}+\hat{U}$ with open boundary conditions. Denoting its eigenstates and eigenenergies by $|n\rangle$ and $E_{n}$, respectively, the time-evolved state is given as follows
\begin{equation}
|\Psi(t)\rangle =  \sum_{n}e^{-iE_{n}t/\hbar}c_{n}|n\rangle,
\label{Eq12}
\end{equation}
where $c_{n}= \langle n|\Psi(0) \rangle$ are the overlaps of the initial state with the Hamiltonian eigenstates. We characterize the dynamics through the two-particle correlation function $C_{ij}=\langle\hat{n}_{i,\uparrow}\hat{n}_{j,\downarrow}\rangle$, where expectation values are evaluated in the time-evolved state~\eqref{Eq12}. In particular, we introduce the total on-site probability and the normalized on-site distribution as follows
\begin{equation}
\Gamma(t)=\sum_j \langle\hat{n}_{j,\uparrow}\hat{n}_{j,\downarrow}\rangle,
\qquad
\Gamma_j(t)=\frac{\langle\hat{n}_{j,\uparrow}\hat{n}_{j,\downarrow}\rangle}{\Gamma(t)}.
\label{}
\end{equation}
Values $\Gamma\simeq 1$ indicate predominantly on-site configurations, as expected for a tightly bound pair. Conversely, $\Gamma\ll 1$ signals primarily weight at finite interparticle separations. Although bound states also contribute to off-diagonal correlations, these are mostly dominated by scattering states. Thus, $\Gamma$ provides an experimentally accessible estimate of the dimer persistence over time. Correspondingly, the normalized distribution $\Gamma_{j}$ resolves the spatial propagation of the tightly bound state and therefore provides a convenient measure of whether the pair remains localized or spreads across the lattice.

Figures~\ref{Fig5}(a) and \ref{Fig5}(b) illustrate the time evolution of $\Gamma_{j}$ for the hopping ratios $J_{\uparrow}/J_{\downarrow}=0.6$ and $J_{\uparrow}/J_{\downarrow}=1.5$, respectively. We consider $(\Delta/J_{\downarrow},U/J_{\downarrow})=(0.4,-6)$ and an evolution time up to $J_{\downarrow}t/\hbar=35$. For the smaller hopping ratio, chosen within the localized regime, the distribution remains sharply localized near the initial site. Increasing the hopping ratio to $J_{\uparrow}/J_{\downarrow}=1.5$ places the pair in the extended regime and produces, after a short transient, a progressively broadening distribution. The red dashed lines in each panel correspond to the ballistic fronts predicted for a tightly bound pair in the absence of quasiperiodicity $\pm 4J_{\uparrow}J_{\downarrow}t/(\hbar|U|)$. These curves show that spreading takes place over the same time window in the unmodulated lattice. Thus, the confinement in Fig.~\ref{Fig5}(a) reflects pair localization due to the quasiperiodic modulation rather than slow pair dynamics. The contrast is more evident in Fig.~\ref{Fig5}(c), which compares the spatial profiles at $J_{\downarrow}t/\hbar=20$. The hopping ratio $J_{\uparrow}/J_{\downarrow}=0.6$ yields a pronounced central peak, whereas $J_{\uparrow}/J_{\downarrow}=1.5$ produces a broader profile.

Figure~\ref{Fig5}(d) illustrates the corresponding time evolution of the total on-site probability $\Gamma(t)$. Starting from $\Gamma(0)=1$, both curves exhibit a rapid initial decrease, associated with the scattering contribution that separates rapidly, followed by an approximately stationary regime with weak residual oscillations. The depletion becomes progressively stronger with increasing $J_{\uparrow}/J_{\downarrow}$. Nevertheless, the on-site probability remains as large as $\Gamma\simeq 0.75$ for $J_{\uparrow}/J_{\downarrow}=1.5$, demonstrating that the observed spatial expansion occurs while a substantial correlated-pair component is retained.

\begin{figure}[t]
\centering
\includegraphics[width=0.99\columnwidth]{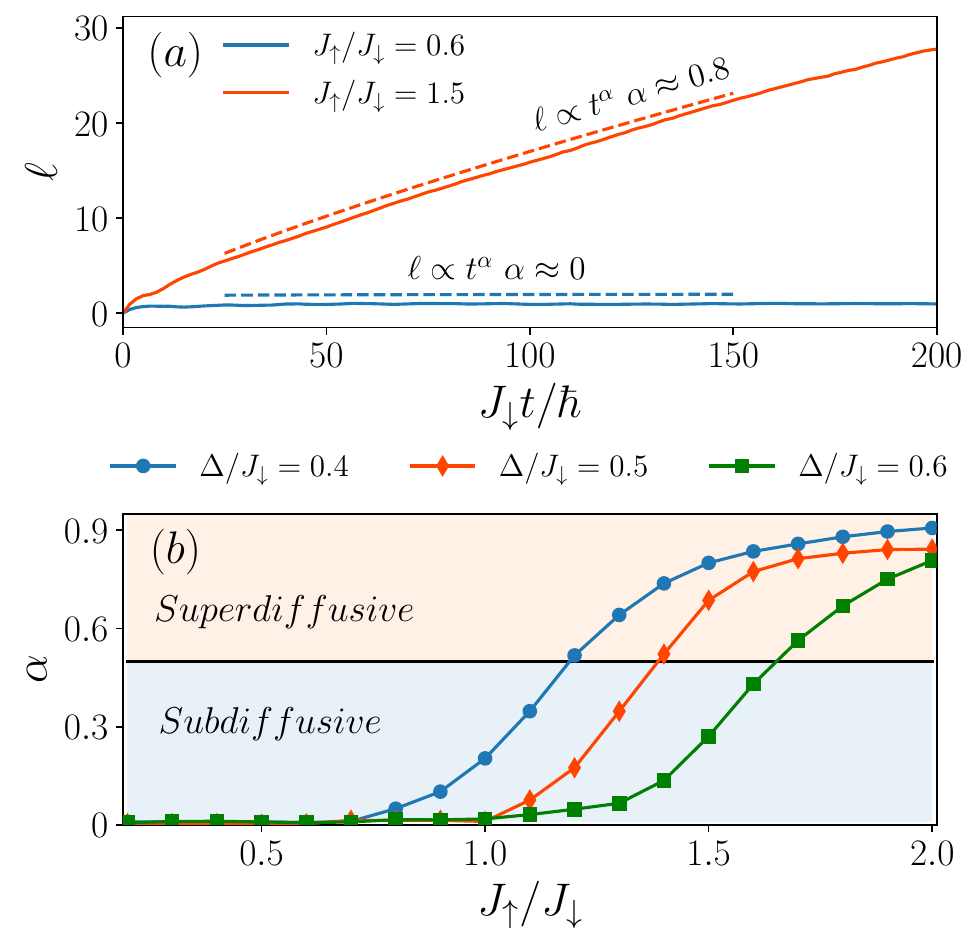}
\caption{(a) Time evolution of the root-mean-square width $\ell(t)$ of the normalized on-site distribution for $J_{\uparrow}/J_{\downarrow}=0.6$ and $J_{\uparrow}/J_{\downarrow}=1.5$, at $(\Delta/J_{\downarrow},U/J_{\downarrow})=(0.4,-6)$. Dashed lines illustrate the power-law scaling $\ell(t)\propto t^\alpha$. (b) Effective dynamical spreading exponent $\alpha$ as a function of $J_{\uparrow}/J_{\downarrow}$ for several values of $\Delta/J_{\downarrow}$. 
The horizontal black line indicates the diffusive regime $\alpha=0.5$. Results are obtained for $U/J_{\downarrow}=-6$ and a lattice of $L=101$ sites. We averaged over $50$ random realizations of the phase $\phi$.\label{Figure6}}
\end{figure}

To quantify the spreading of the on-site distribution $\Gamma_{j}$ at longer times, we compute its root-mean-square (RMS) width 
\begin{equation}
\ell(t) = \left[\sum_{j}x_j^2\Gamma_{j}(t)-\left(\sum_{j}x_{j}\Gamma_{j}(t)\right)^2\right]^{1/2},
\label{}
\end{equation}
where $x_{j}=j-j_{0}$ with $j_{0}=L/2$ the initial site. Figure~\ref{Figure6}(a) compares $\ell(t)$ for the same parameters as in Fig.~\ref{Fig5}. For $J_{\uparrow}/J_{\downarrow}=0.6$, $\ell(t)$ remains small and approximately constant. In contrast, for  $J_{\uparrow}/J_{\downarrow}=1.5$, the RMS width grows in time, demonstrating sustained propagation of the on-site pair component.

We characterize the growth of the RMS width using the power-law fit $\ell(t)\propto t^{\alpha}$, where $\alpha$ is the effective dynamical spreading exponent within the investigated time window. The fitting procedure is performed over intermediate-time intervals, excluding the initial transient and, where relevant, boundary effects. The resulting exponents are $\alpha\approx0$ for $J_{\uparrow}/J_{\downarrow}=0.6$ and $\alpha\approx0.8$ for $J_{\uparrow}/J_{\downarrow}=1.5$. The former is consistent with localization, whereas the latter describes superdiffusive spreading.

Figure~\ref{Figure6}(b) shows the effective dynamical spreading exponent $\alpha$ as a function of the hopping ratio $J_{\uparrow}/J_{\downarrow}$ for several quasiperiodic strengths. At hopping ratios within the localized regime, $\alpha\approx 0$, consistent with suppressed pair transport. As the hopping ratio increases, $\alpha$ rises through subdiffusive ($0<\alpha<0.5$) and superdiffusive ($0.5<\alpha<1$) values, approaching ballistic ($\alpha=1$) behavior for the weakest modulation shown. Increasing $\Delta/J_{\downarrow}$ shifts the onset of spreading toward larger hopping ratios, in agreement with the time-independent analysis.

\begin{figure}[t]
\centering
\includegraphics[width=0.99\columnwidth]{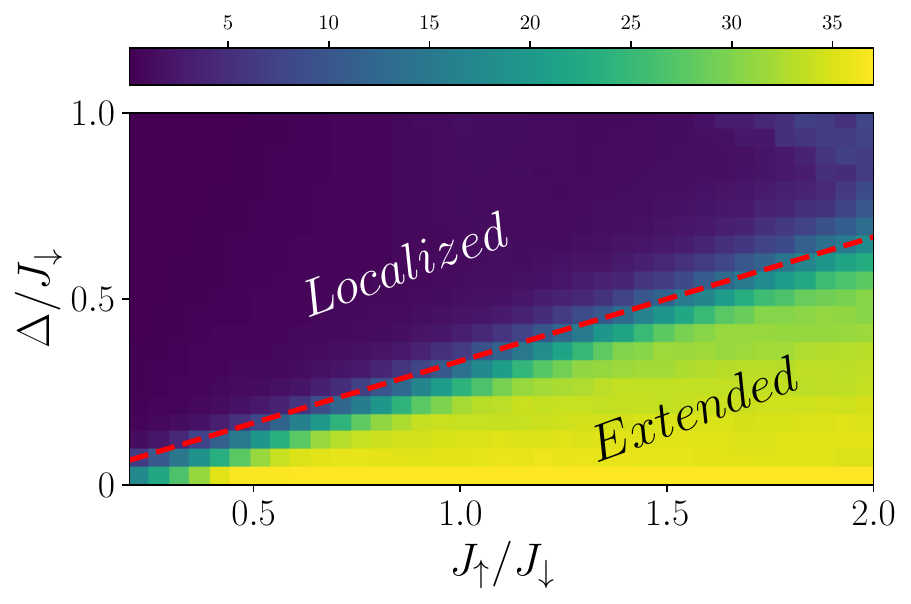}
\caption{Maximum width $\ell_{\max}$ reached during a time evolution up to $J_{\downarrow}t_{max}/\hbar=200$, as a function of the hopping ratio and the quasiperiodic modulation. The red dashed line denotes the strong-coupling prediction $\Delta_{SC}\sim 2J_{\uparrow}J_{\downarrow}/|U|$. Results are obtained for $U/J_{\downarrow}=-6$ and a lattice of $L=101$ sites. We averaged over $50$ random realizations of the phase $\phi$.\label{Figure7}}
\end{figure}

To conclude, we summarize the spatial extent reached during the time evolution through its maximum value $\ell_{max}=\max_{0\leq t\leq t_{max}} \ell(t)$, with $J_{\downarrow}t_{max}/\hbar=200$. Figure~\ref{Figure7} shows $\ell_{max}$ as a function of the hopping ratio and the quasiperiodic strength. A region of $\ell_{\max}\approx a$, associated with localized pair dynamics, is separated from a region of extended propagation $\ell_{\max}\gg a$. The strong-coupling prediction $\Delta_{SC}\sim 2J_{\uparrow}J_{\downarrow}/|U|$, indicated by the red dashed line, reasonably tracks the crossover between localized and extended dynamics.

\section{Conclusions}
We have investigated how unequal hopping amplitudes shape quantum localization of two interacting particles in a one-dimensional quasicrystal. By employing a two-body Green's function approach, we determined the localization phase diagram of the attractively bound dimer ground state and the highest-energy repulsively bound pair. Increasing the hopping amplitude of one constituent at fixed hopping of the other raises the localization threshold, which remains bounded from above by the single-particle Aubry–André critical point of the slower constituent. Effective analytical descriptions account for pair localization in the limits of strong hopping imbalance and strong interactions, while numerical results demonstrate continuous tunability of the pair localization length within the localized regime. The expansion of an on-site pair provides complementary dynamical signatures of delocalization induced by unequal hops. In particular, the effective spreading exponent of the normalized on-site pair distribution reveals a progression from localized to subdiffusive and eventually superdiffusive dynamics as the hopping ratio increases.

Our results establish hopping imbalance as a means of tuning the spatial extent and propagation of interacting pairs in quasiperiodic lattices, with direct relevance to ultracold atoms in optical lattices. An interesting direction for future work is to investigate how coupling components with different kinetic scales modifies their localization properties beyond the two-particle limit, particularly the backaction of mobile particles on a localized component.

\begin{acknowledgments}
G.A.D.-C. acknowledges financial support from UNAM DGAPA PAPIIT Grant No. IA101327. F.M. and U.A.D.-R. acknowledge financial support from UNAM DGAPA PAPIIT Grant No. IN111624. U.A.D.-R. further acknowledges support from the UNAM-DGAPA
postdoctoral scholarship.
\end{acknowledgments}

\appendix
\section{Periodic lattice}
\label{Periodic}

\begin{figure}[t!]
\centering
\includegraphics[width=0.99\columnwidth]{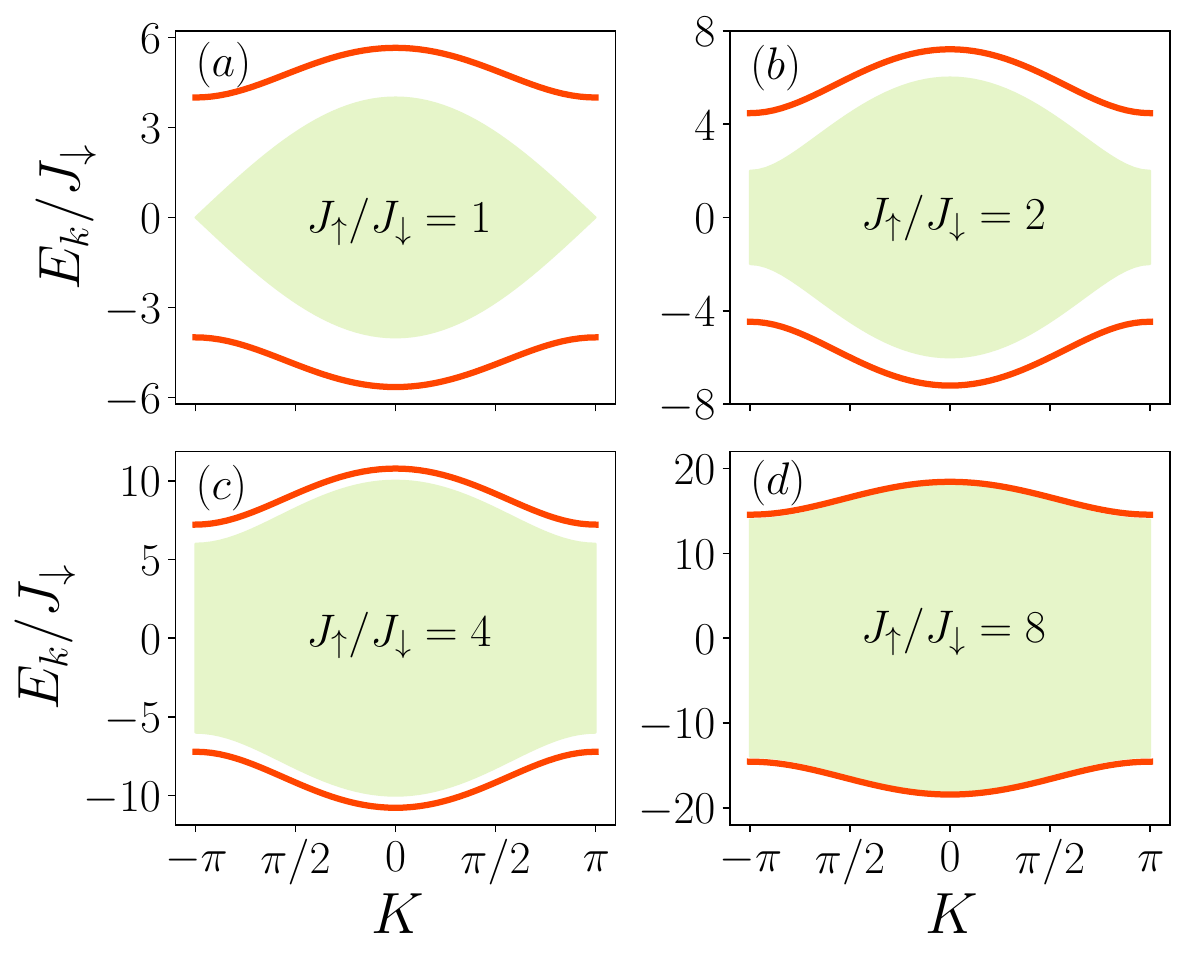}
\caption{(a)-(d) Two-body energy spectrum as a function of the collective quasimomentum $K$ for several hopping ratios $J_{\uparrow}/J_{\downarrow}$. The green shaded regions denote the continuum scattering states, and the red curves are associated with the bound state branches $U/J_{\downarrow}=\pm 4$.\label{FigA1}}
\end{figure}

In this Appendix, we summarize the properties of two interacting particles moving in a translationally invariant lattice ($\Delta=0$) with unequal hopping amplitudes. For further details see Ref.~\cite{PhysRevA.78.033611}. In the noninteracting limit ($U=0$), the two-body spectrum is the sum of the single-particle dispersions $E_{k_{\uparrow},k_{\downarrow}} = \varepsilon_{k_{\uparrow},\uparrow}+\varepsilon_{k_{\downarrow},\downarrow}$, where $\varepsilon_{k_{\sigma},\sigma}=-2J_{\sigma}\cos k_{\sigma}a$ and $a$ is the lattice constant, set to $a=1$ below. Introducing the collective and relative quasimomentum coordinates, $K=k_{\uparrow}+k_{\downarrow}$ and $q=(k_{\uparrow}-k_{\downarrow})/2$, respectively, gives
\begin{equation}
\begin{split}
E_{K,q} =&
-2J_{\uparrow}\cos\left(\frac{K}{2}+q\right)\\
&-2J_{\downarrow}\cos\left(\frac{K}{2}-q\right).
\end{split}
\label{Aeq1}
\end{equation}
Defining the average and half-difference hopping amplitudes,
$J_{\Sigma}=(J_{\uparrow}+J_{\downarrow})/2$ and
$J_{\Delta}=(J_{\uparrow}-J_{\downarrow})/2$, we rewrite Eq.~\eqref{Aeq1} as
\begin{equation}
E_{K,q} = W_{K}\cos(T_{K}+q),
\label{Aeq2}
\end{equation}
where 
\begin{equation}
W_{K} = -4\sqrt{J_{\Sigma}^{2}\cos^{2}K/2+J_{\Delta}^{2}\sin^{2}K/2}, 
\label{Aeq3}
\end{equation}
sets the continuum half-width $|W_K|$, and
\begin{equation}
T_{K} = \sin^{-1}\frac{4J_{\Delta}\sin (K/2)}{|E_{K}|}.
\label{Aeq4}
\end{equation}
is a shift of the relative quasimomentum. Figures~\ref{FigA1}(a)-\ref{FigA1}(d), show the scattering continuum as a function of $K$ for several hopping ratios $J_{\uparrow}/J_{\downarrow}$. For equal hopping amplitudes ($J_{\uparrow}/J_{\downarrow}=1$), $E_{K,q}$ vanishes at the Brillouin-zone boundaries, $K=\pm\pi$. In contrast, unequal hopping amplitudes yield a finite width of $2|W_{\pm\pi}|=4|J_{\uparrow}-J_{\downarrow}|$.

In the presence of on-site interactions, translational invariance allows the two-body wave function to be written as $\Psi(j_{\uparrow},j_{\downarrow})=e^{iKR}\Phi_{K}(r)$, where $R=(j_{\uparrow}+j_{\downarrow})/2$ and $r=j_{\uparrow}-j_{\downarrow}$ are the collective and relative coordinates, respectively. Substituting this ansatz into Eq.~\eqref{Eq5} yields the bound state condition
$UG_{0,K}(E,0)=1$, with
\begin{equation}
G_{0,K}(E,r) = \frac{1}{2\pi}\int_{-\pi}^{\pi} dq \ \frac{e^{iqr}}{E-E_{Kq}+i0^{+}},
\label{Aeq5}
\end{equation}
the noninteracting Green's function for relative motion at fixed $K$. Solving this condition gives~\cite{PhysRevA.78.033611}
\begin{equation}
E_{dimer,K}=\pm\sqrt{U^2+4J_\uparrow^2+4J_\downarrow^2
+8J_\uparrow J_\downarrow\cos K}
\label{Aeq6}
\end{equation}
In Figs.~\ref{FigA1}(a)-\ref{FigA1}(d), we show the bound pair energies as a function of the collective quasi-momentum coordinate $K$ for $U/J_{\downarrow}=\pm 4$. 
\begin{figure}[t!]
\centering
\includegraphics[width=0.99\columnwidth]{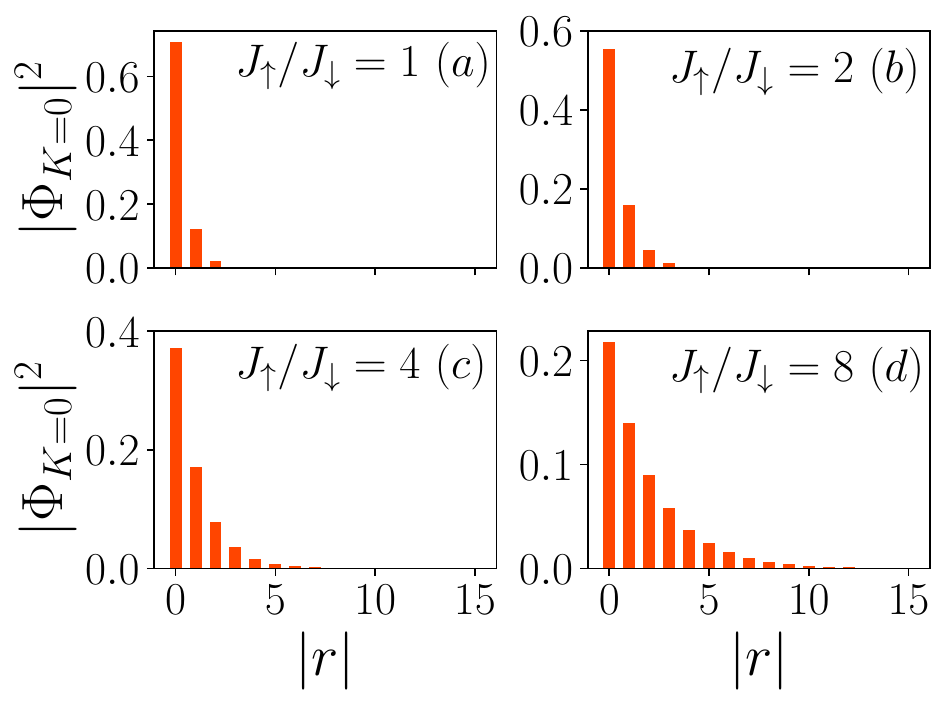}
\caption{(a)-(d) Square modulus of the zero quasi-momentum wave function $|\Phi_{K=0}(r)|^{2}$ as a function of the interparticle separation $|r|$ for several hopping ratios $J_{\uparrow}/J_{\downarrow}$. We consider a fixed interaction strength of $U/J_{\downarrow}=-4$.\label{FigA2}}
\end{figure}

Up to a normalization factor, the relative bound state wave function is proportional to the Green’s function evaluated at the bound-state energy, $\Phi_{K}(r)\propto G_{0,K}(E,r)$. Figures~\ref{FigA2}(a)-\ref{FigA2}(d) illustrate the corresponding probability distribution at $K=0$ for $U/J_{\downarrow}=-4$ and several hopping ratios. Increasing $J_{\uparrow}$ at fixed $J_{\downarrow}$ and $U$ broadens the distribution, indicating an increase in the characteristic interparticle separation. This broadening accompanies the reduction of the binding energy relative to the continuum edge at $K=0$.

\section{Effective dimer model for strongly imbalanced hopping}
\label{Quasiperiodic}

In this Appendix, we derive the analytical expressions for the effective dimer hopping and quasiperiodic modulation amplitude in the strongly imbalanced regime, $J_\uparrow\gg J_\downarrow$. We consider attractive interactions $U<0$, the corresponding results for $U>0$ are obtained in a straightforward manner. For a translationally invariant lattice ($\Delta=0$), the bound pair dispersion in Eq.~\eqref{Aeq6} can be written as
\begin{equation}
E_{dimer,K}=-\sqrt{U^2+4J_\uparrow^2}
\left[
1+\frac{4J_\downarrow^2+
8J_\uparrow J_\downarrow\cos K}
{U^2+4J_\uparrow^2}
\right]^{1/2}.
\end{equation}
In the strongly imbalanced regime $(J_{\uparrow},|U|\gg J_{\downarrow})$,
\begin{equation}
\frac{4J_\downarrow^2+
8J_\uparrow J_\downarrow\cos K}
{U^2+4J_\uparrow^2}\ll 1.   
\end{equation}
We expand the square root to first order in $J_\downarrow$, yielding
\begin{equation}
E_K\simeq-\sqrt{U^2+4J_\uparrow^2}
-\frac{4J_\uparrow J_\downarrow}
{\sqrt{U^2+4J_\uparrow^2}}\cos K.
\end{equation}
The resulting dispersion has the nearest-neighbor tight-binding form $E_K\simeq E_0-2J_{\mathrm dimer}\cos K$, with
\begin{equation}
E_0=-\sqrt{U^2+4J_\uparrow^2},
\qquad
J_{\mathrm dimer}=
\frac{2J_\uparrow J_\downarrow}
{\sqrt{U^2+4J_\uparrow^2}}.
\end{equation}
Here, $E_{0}$ is the bound state energy for an immobile $\downarrow$ particle, whereas $J_{\mathrm dimer}$ is the effective dimer hopping. This expansion retains the full dependence on $|U|$ and $J_\uparrow$. For $|U|\gg J_\uparrow\gg J_{\downarrow}$, it recovers the strong-coupling result $J_{\mathrm dimer}\simeq2J_\uparrow J_\downarrow/|U|$. Conversely, for $J_\uparrow\gg|U|\gg J_{\downarrow}$, the effective hopping approaches $J_{\mathrm dimer}\to J_\downarrow$, indicating that pair motion is limited by the slower constituent.

To derive the effective quasiperiodic modulation, we first determine the bound state profile at $\Delta=0$ in the immobile-particle limit, $J_\downarrow=0$. Fixing the $\downarrow$ particle at site $j_{\downarrow}$, the Hamiltonian of the mobile $\uparrow$ particle in the relative coordinate $r=j_\uparrow-j_{\downarrow}$ is
\begin{equation}
\hat{H}_\uparrow
=
-J_\uparrow\sum_r
\left(|r+1\rangle\langle r|+\mathrm{h.c.}\right)
+U|0\rangle\langle0|,
\end{equation}
For $U<0$, we seek a solution of the form $
\chi(r)=A e^{-\kappa|r|}$ with $\kappa>0$. For $r\neq0$, the Schrödinger equation gives $E_0=-2J_\uparrow\cosh\kappa$, whereas at $r=0$ it yields $E_0=U-2J_\uparrow e^{-\kappa}$. Combining these relations, we obtain
\begin{equation}
\sinh\kappa=\frac{|U|}{2J_\uparrow}.
\label{}
\end{equation}
Normalization then gives the relative-coordinate probability distribution $|\chi(r)|^2=\tanh\kappa\,e^{-2\kappa|r|}$. For $J_\uparrow\gg|U|$, the inverse decay length becomes $\kappa\simeq|U|/(2J_\uparrow)$.

We now include the quasiperiodic potential
\begin{equation}
V(j_{\uparrow},j_{\downarrow})=\Delta\left[\cos(2\pi\beta j_{\downarrow}+\phi)+\cos(2\pi\beta j_{\uparrow}+\phi)\right].
\label{}
\end{equation}
For the slow particle fixed at $j_\downarrow$, averaging over the unperturbed bound state density ($\sum_r |\chi(r)|^{2}=1$) gives the potential energy shift to first order in $\Delta$
\begin{equation}
\begin{split}
\overline V(j_{\downarrow}) &=
\Delta\cos(2\pi\beta j_{\downarrow}+\phi)\\
&+
\Delta\sum_r |\chi(r)|^{2}\cos[2\pi\beta(r+j_{\downarrow})+\phi].
\end{split}
\end{equation}
Using the identity
\begin{equation}
\begin{split}
\cos[2\pi\beta(r+j_{\downarrow})+\phi] &= \cos(2\pi\beta j_{\downarrow}+\phi)\cos(2\pi\beta r)\\
& -\sin(2\pi\beta j_{\downarrow}+\phi)\sin(2\pi\beta r),
\label{}
\end{split}
\end{equation}
together with the inversion symmetry $|\chi(r)|=|\chi(-r)|$, the contribution odd in $r$ vanishes. The averaged potential therefore retains the original quasiperiodic form
\begin{equation}
\overline V(j_{\downarrow})=\Delta[1+F]\cos(2\pi\beta j_{\downarrow}+\phi),    
\end{equation}
where  $F =\sum_{r}|\chi(r)|^2\cos(2\pi\beta r)$ is the form factor of the relative coordinate probability. Substituting the exponential density and evaluating the geometric series yields
\begin{equation}
F
=
\frac{\sinh^2\kappa}
{\sinh^2\kappa+\sin^2(\pi\beta)}
= \frac{|U|^2}{|U|^2+4J_\uparrow^2\sin^2(\pi\beta)},
\end{equation}
where we have used $\sinh\kappa=|U|/(2J_\uparrow)$. Thus, to leading order in the quasiperiodic modulation, the effective dimer potential has the amplitude
\begin{equation}
\Delta_{\mathrm{eff}}
=
\Delta\left[
1+
\frac{|U|^2}{|U|^2+4J_\uparrow^2\sin^2(\pi\beta)}
\right].    
\end{equation}
For $J_\uparrow\gg|U|\gg J_\downarrow$, spatial averaging over the extended bound-state profile suppresses the contribution of the mobile particle, yielding $\Delta_{\mathrm{dimer}}\to\Delta$. Conversely, in the strong-coupling regime, $|U|\gg J_\uparrow\gg J_\downarrow$, the two particles predominantly occupy the same site, recovering $\Delta_{\mathrm{dimer}}\to2\Delta$. 

\bibliography{bib-twoAA}
\end{document}